# Low-Frequency 1/*f* Noise in $MoS_2$ Thin-Film Transistors: Comparison of Single and Multilayer Structures

S. L. Rumyantsev[1,2], C. Jiang[1,2], R. Samnakay[3,4] M.S. Shur[1] and A.A. Balandin[3,4]

[1]Department of Electrical, Computer, and Systems Engineering, Center for Integrated Electronics, Rensselaer Polytechnic Institute, Troy, New York 12180, USA

[2]Ioffe Physical-Technical Institute, St. Petersburg 194021, Russia

[3]Nano-Device Laboratory (NDL), Department of Electrical and Computer Engineering, University of California – Riverside, Riverside, California 92521 USA

[4]Phonon Optimized Engineered Materials (POEM) Center, Materials Science and Engineering Program, University of California – Riverside, Riverside, California 92521 USA

## Abstract

We report on the transport and low-frequency noise measurements of $MoS_2$ thin-film transistors with “thin” (2-3 atomic layers) and “thick” (15-18 atomic layers) channels. The back-gated transistors made with the relatively thick $MoS_2$ channels have advantages of the higher electron mobility and lower noise level. The normalized noise spectral density of the low-frequency 1/*f* noise in “thick” $MoS_2$ transistors is of the same level as that in graphene. The $MoS_2$ transistors with the atomically thin channels have substantially higher noise levels. It was established that, unlike in graphene devices, the noise characteristics of $MoS_2$ transistors with “thick” channels (15-18 atomic planes) could be described by the McWhorter model. Our results indicate that the channel thickness optimization is crucial for practical applications of $MoS_2$ thin-film transistors.

Atomically thin two-dimensional (2D) materials – also referred to *as van der Waals* materials – are attracting attention for electronic, sensing and optical applications. So far, the most explored materials among them are graphene and one to several atomic layers $MoS_2$ [1-3]. While single-layer graphene is a zero band gap material, single-layer $MoS_2$ shows a direct band gap of $E_g$~1.9 eV. As the number of layers increases, the band gap of $MoS_2$ decreases until it reaches its bulk value of 1.3 eV [4-6]. With the exception of single-layer, the thin films of $MoS_2$ of all other thicknesses, including bulk, are indirect band gap semiconductors. A relatively large band gap of atomic layers of $MoS_2$ is one of its main advantages over graphene, since it makes MoS2 suitable for transistor applications. Unlike graphene devices, $MoS_2$ thin-film transistors (TFTs) demonstrate very high on-off ratios of up to $10^8$ [1-6]. A relatively high mobility of up to a few hundreds of $cm^2/Vs$ and high thermal conductivity of $MoS_2$ thin films might make $MoS_2$ transistors competitive with $\alpha$-Si and poly-Si TFTs [1-7].

An interesting feature of the 2D materials is the dependence of their properties on the number of the atomic layers. A proper selection of the thickness, i.e. number of atomic layers in $MoS_2$ films for specific device applications is one of the important issues in the development of the *van der Waals* materials technology. While single layer (three atomic planes S-Mo-S) devices are attractive for sensing and optical applications because of their ultimate surface-to-volume ratio [8-11], multi-layer $MoS_2$ films might have an advantage for electronics applications. Compared to graphene and single layer MoS2 devices, the multilayer $MoS_2$ TFTs have a higher stability, lesser sensitivity to the environment and higher electron mobility. In particular, as shown in Ref. [12], the electron mobility in $MoS_2$ films has maximum at thickness H=10 nm, which corresponds to approximately 15-16 layers of the material.

An important parameter for electronics and sensing applications of transistors is the level of the low-frequency electronic noise. In various types of sensors, the low frequency noise sets the sensitivity and selectivity limits ([13] and references therein). For high-frequency communication applications, the low-frequency noise is up converted to the phase noise, thus limiting the performance of every microwave and terahertz device. The studies of the low-frequency noise in $MoS_2$ transistors have already been reported in a number of publications [14-20]. The experimental results have been interpreted in the framework of either McWhorter number-of-carriers model [14-16], unified model incorporating carrier-number fluctuation and correlated mobility fluctuations, [17-18] or expressed using the empirical Hooge formula [19]. In the majority of these publications, it was found that the normalized noise spectral density of the drain current, $S_I/I^2$, significantly decreased with the increasing gate voltage, $V_g$. One exception was the data presented in Ref. [20] showing a relatively weak gate voltage dependence with the maximum at a certain gate voltage (similar to such a dependence often found in graphene).

The decrease of the noise spectral density of the drain current with the gate voltage in $MoS_2$ TFT is similar to that for conventional Si metal-oxide-semiconductor field-effect transistors (MOSFETs). In conventional MOSFETs, in the majority of cases, the low-frequency noise complies with the McWhorter model [21, 22], which predicts the decrease of $S_I/I^2$ as $1/(V_g-V_t)^2$, where $V_t$ is the threshold voltage. This makes noise properties of $MoS_2$ TFT qualitatively similar to those in MOSFETs and different from those in graphene devices. In graphene, several shapes of the gate voltage dependence of the noise spectral density have been reported. The noise can slightly increase with $(V_g-V_t)$ (demonstrating a "V" shape dependence), slightly decrease or follow the so-called "M" shape dependence, when the noise first increases with increasing $(V_g-V_t)$ and then goes down [23-25]. However, in general all these dependences are weak. Typically, the noise in graphene transistors normalized to the channel area is within the range $10^{-8}$-$10^{-7}$ $\mu m^2$/Hz [26]. In comparison, in Si MOSFETs noise changes by many orders of magnitude decreasing with the gate voltages as $S_I/I^2 \sim 1/(V_g-V_t)^2$, in accordance with the McWhorter model.

In the present work, we report on the low-frequency noise in "thin" (2-3 layers) and "thick" (15-18 layers) back-gated $MoS_2$ TFTs. The focus of this study is on comparison of the noise level in these two types of $MoS_2$ TFTs with that benchmarking it against noise data published for back-gated graphene devices. Thin films of $MoS_2$ were prepared by a standard exfoliation method and placed on $Si/SiO_2$ substrates. Details of material preparation can be found in Refs. [28-29] and references therein. The thickness and quality of thin films were determined with the atomic force microscopy (AFM) and micro-Raman spectroscopy. The drain and source Ti/Au (10-nm / 100-nm) contacts were fabricated using the electron-beam lithography (LEO SUPRA 55) for patterning of the source and drain electrodes and the electron-beam evaporation (Temescal BJD-1800) for metal deposition. The heavily doped $Si/SiO_2$ wafer served as a back gate. The channel length, L, and width, W, varied within the range from 1.3 μm to 6 μm. Figure 1 shows a schematic and scanning electron microscopy (SEM) image of a typical fabricated $MoS_2$ TFT.

Figure 2 presents the transfer current-voltage characteristics of $MoS_2$ TFTs with different channel thicknesses. The lateral dimensions of all examined devises are similar. The characteristics for graphene devices are also shown for the comparison. Some specific advantages and disadvantages of these three systems are already seen from these current-voltage characteristics. While graphene device are characterized by a small on-off ratio, $I_{on}/I_{off}$, the $MoS_2$ TFTs demonstrate $I_{on}/I_{off}>10^5$. Graphene devices have substantially higher current levels than the "thin" (2-3 layers) $MoS_2$ TFTs owing to a much higher carriers mobility in graphene. The drain current in the "thick" (15-18 layers) $MoS_2$ TFTs is higher than that in "thin" MoS2 TFTs and approaches the typical values for graphene devices. The subthreshold voltage slope in the "thick" $MoS_2$ TFTs is smaller than that in the "thin" $MoS_2$ TFTs, i.e. a higher gate voltage swing is required to switch the "thick" $MoS_2$ TFTs. Plotting the drain-to-source resistance, $R_{ds}$, vs. $1/(V_g-V_{th})$, and extrapolating this dependence to zero yielded the estimate for the total contact resistance, which is negligible for the "thick" $MoS_2$ devices compared to the channel resistance.

The field-effect mobility, $\mu_{FE}$, in $MoS_2$ TFTs was calculated as

$$\mu_{FE} = \frac{g_m L}{C_{ox} V_d W}. \qquad (1)$$

Here , $g_m$ is the transconductance, $C_{OX} = \varepsilon_o \varepsilon_r / d = 1.15 \times 10^{-4}$ (F/m$^2$) is the oxide capacitance, $\varepsilon_o$ is the dielectric permittivity of free space, $\varepsilon_r$ is the dielectric constant, and $d$ is the oxide thickness. We used $\varepsilon_r$=3.9 and $d$=300 nm for the $SiO_2$ layer. We found the field-effect mobility within the range $\mu_{FE}$=0.5 – 8 cm$^2$/Vs for "thin" $MoS_2$ devices (thickness H=2-3 atomic layers) and $\mu_{FE}$=20 – 80 cm$^2$/Vs for "thick" $MoS_2$ devices (H=15-18 atomic layers). The fact that the mobility in "thick" $MoS_2$ TFTs is on the order of magnitude higher than that in "thin" devices is in agreement with the results reported in Ref. [12]. Overall, the mobility values in our "thin" and "thick" $MoS_2$ TFTs are typical for such back-gated devices and in agreement with the previously reported values [12].

The low-frequency noise was measured under ambient conditions at room temperature (RT) in the linear regime at the drain voltage $V_d$=50 – 100 mV. The low-frequency noise was of the $1/f^{\alpha}$ type without any generation-recombination bulges. The extracted exponent is $\alpha$=0.75-1.25. Figure 3 compares the gate-voltage dependence of 1/f noise spectral density normalized to the device area for four types of devices: "thin" $MoS_2$ TFTs (H=2-3 atomic layers), "thick" $MoS_2$ TFTs (H=15-18 atomic layers), single layer graphene field-effect transistors (FETs), and thick layer graphene. As seen, "thick" $MoS_2$ TFTs have smaller noise levels, comparable to that in graphene devices. While at small gate voltage, the noise levels in "thin" and "thick" $MoS_2$ films are comparable, at $(V_g-V_t)>$ 10 V, the noise level in "thick" $MoS_2$ films is much smaller. We attribute the noise increase in "thin" $MoS_2$ films with the gate voltage to the contribution of the contact noise and effects of the fast aging of these devices.

The McWhorter model allows estimating the trap density responsible for the low-frequency noise [21] from the drain current fluctuation density, $S_I/I^2$, or from the equivalent voltage fluctuations, $S_V=S_I/g_m^2$, where $g_m$ is the transconductance. In the latter case, the trap density can be estimated as [22]

$$N_t = S_v \frac{\gamma f W L C_{ox}^2}{kTq^2}, \tag{2}$$

where $k_B$ is the Boltzmann constant, T is absolute temperature, $\gamma$ is the tunneling parameter conventionally assumed to be $\gamma=10^8$ cm$^{-1}$, and $N_t$ is the trap density. The advantage of using this approach in comparison with the trap density estimation from the drain current fluctuations is that it is independent of the threshold voltage, $V_t$, which can have high uncertainty is some cases. Our estimate for the TFTs with the "thick" $MoS_2$ channel yields $N_t \approx 10^{18}$ cm$^{-3}$eV$^{-1}$ (for the smallest noise device with the characteristics shown in Fig.3). This value is about one order-of-magnitude smaller than that for as fabricated $MoS_2$ TFTs with H=3 atomic planes described in Ref. [14]. The interface trap density for $MoS_2$ TFTs with the channels of a similar thickness was found to be $N_{it}=(5.5 - 7.2)\times10^{10}$ cm$^{-2}$eV$^{-1}$ [15]. An estimate of the interface trap density depends on the estimation of the characteristic tunneling depth in $SiO_2$. Using the same approach as in Ref. [15], we obtain $N_{it}\approx10^{10}$ cm$^{-2}$eV$^{-1}$, which is only slightly smaller than the value in Ref. [15].

In conclusion, we confirmed that the low-frequency noise in $MoS_2$ TFTs with "thick" channels follows McWhorter model in contrast to that in graphene. It was established that the low-frequency noise level in $MoS_2$ TFTs depends strongly on the thickness of the device channel. The $MoS_2$ devices with "thin" channels (H=2-3 atomic layers) have a relatively high noise level compared to the devices with "thick" channels (H=15-18 atomic layers), whose noise level is of the same order of magnitude as that in graphene devices. Other characteristics of "thick" channel $MoS_2$ TFTs such as high electron mobility, low contact resistance and high stability constitute additional advantages of this type of structures for electronic applications compared to the $MoS_2$ channels with the thickness of 1-3 atomic planes.

***Acknowledgments***

This work was supported, in part, by the Semiconductor Research Corporation (SRC) and Defense Advanced Research Project Agency (DARPA) through STARnet Center for Function Accelerated nanoMaterial Engineering (FAME). AAB also acknowledges funding from the National Science Foundation (NSF) for the project Graphene Circuits for Analog, Mixed-Signal, and RF Applications (NSF CCF-1217382). SLR acknowledges partial support from the Russian Fund for Basic Research (RFBR). The work at RPI was supported by NSF under the auspices of the EAGER program. The authors thank Professor M. Levinshtein (Ioffe Institute, St. Petersburg, Russia), for helpful discussions.

**FUGURE CAPTIONS**

**Figure 1:** Schematic of the back-gated $MoS_2$ thin-film transistor (a). Scanning electron microscopy image of a typical $MoS_2$ TFT (b). The pseudo-colors are used for clarity: blue is $MoS_2$ channel while yellow is metal contacts.

**Figure 2:** Transfer current-voltage characteristics of graphene devices, $MoS_2$ TFT with "thick" channel (thickness H=15-18 atomic layers) and $MoS_2$ TFT with "thin" channel (H=2-3 atomic planes). The data for "thin" channel devices are shown for the pristine and aged devices.

**Figure 3:** Right panel: Comparison of the noise spectral density versus gate voltage dependences for $MoS_2$ TFTs with "thick" channel (H=15-18 atomic layers), $MoS_2$ TFTs with "thin" channel (H=2-3 atomic planes), single and bi-layer graphene devices (SLG, BLG), and multi-layer graphene devices. The data for graphene devices are taken from Ref. [26, 27]. Black diamond symbols show the data for as fabricated "thin" $MoS_2$ transistor. Other symbols for "thin" $MoS_2$ transistor represent noise data for several devices with different stages of aging ranging from 2 days to several weeks. The "thick" channel transistors demonstrated good stability in current voltage characteristics and noise behavior over at least one month. The data for three TFTs are shown. Left panel: noise ranges for the studied devices.

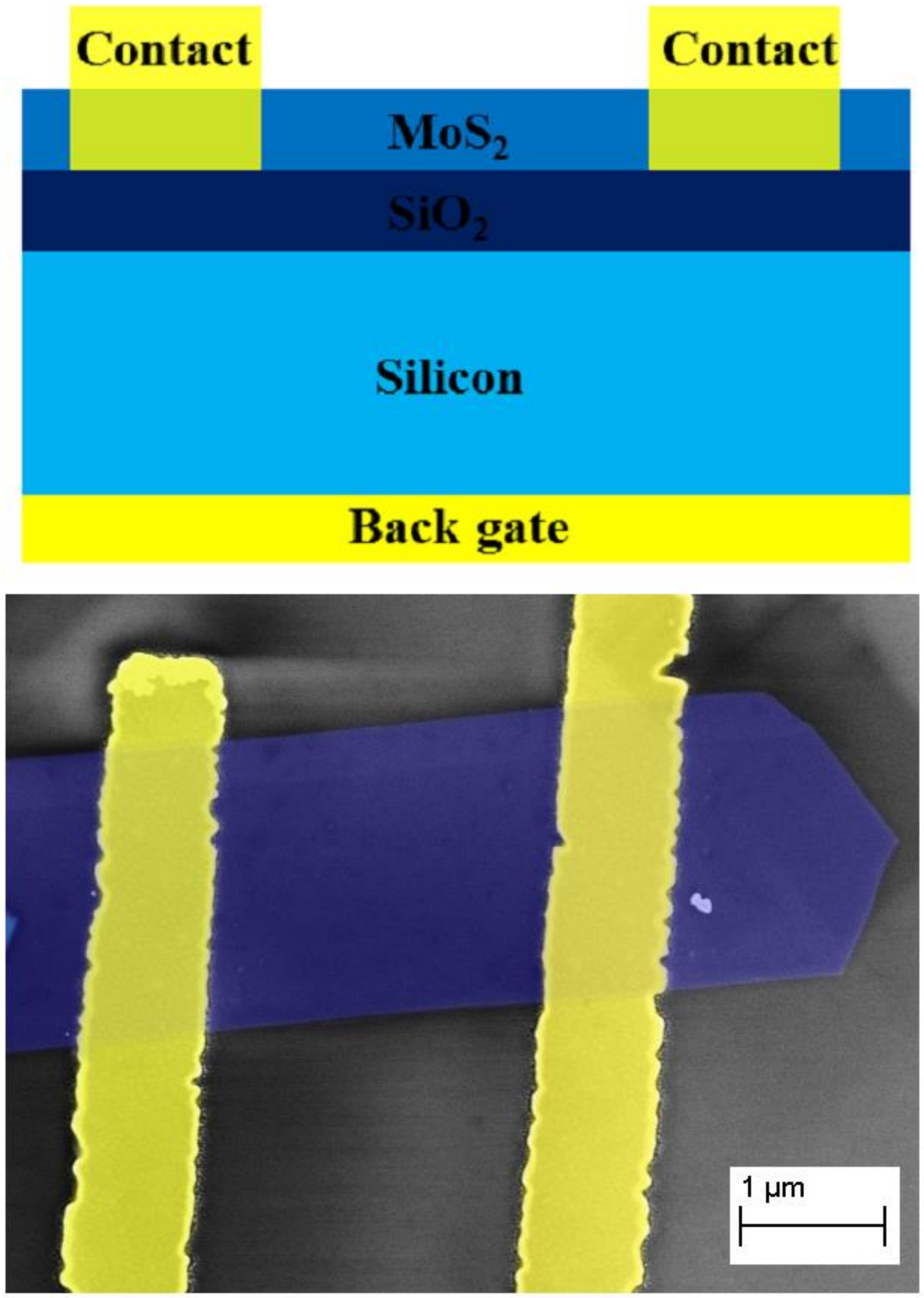

Figure 1

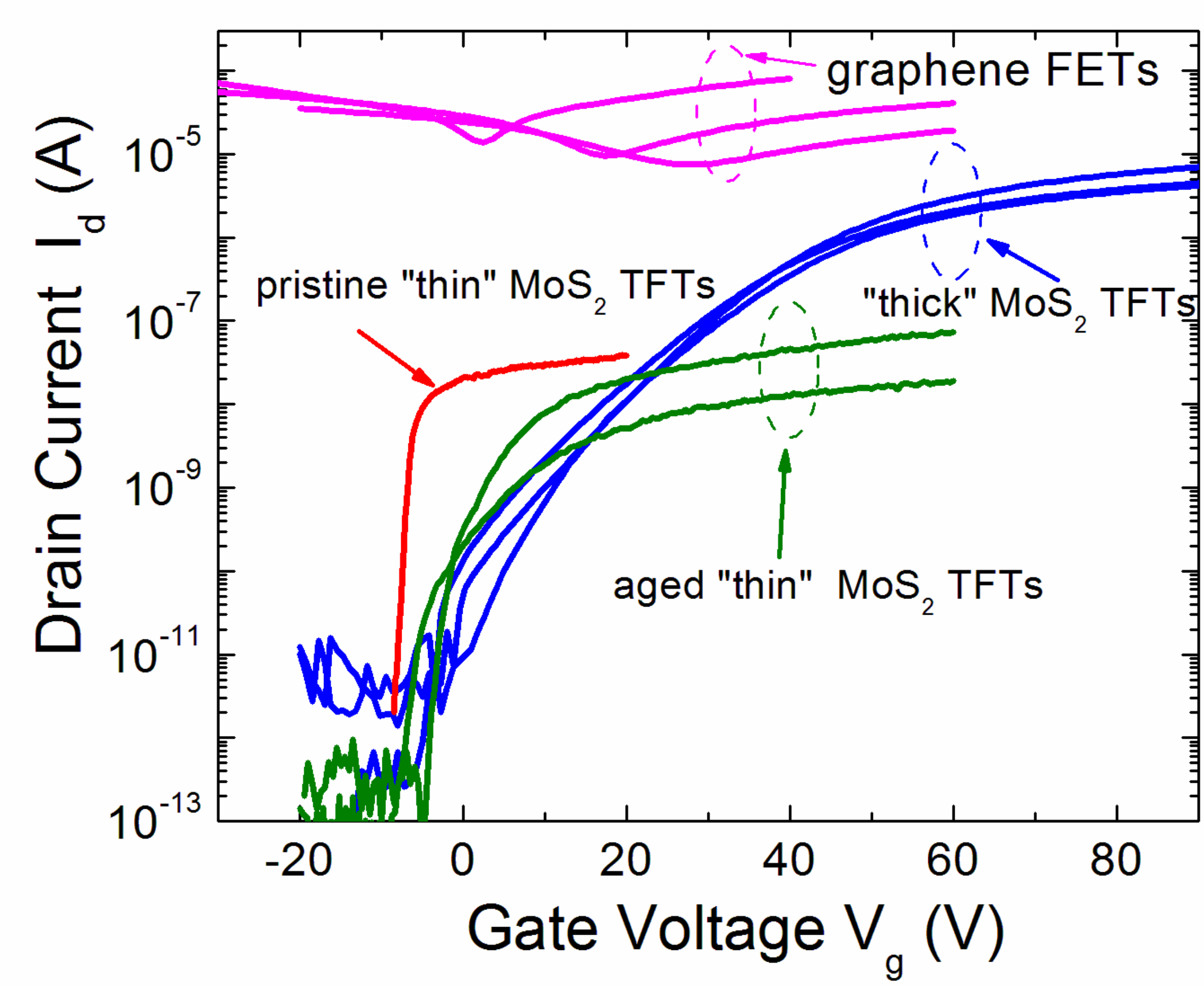
graphene FETs
"thick" MoS2 TFTs
pristine "thin" MoS2 TFTs
aged "thin" MoS2 TFTs
Drain Current I_d (A)
10^-5
10^-7
10^-9
10^-11
10^-13
-20
0
20
40
60
80
Gate Voltage V_g (V)

Figure 2

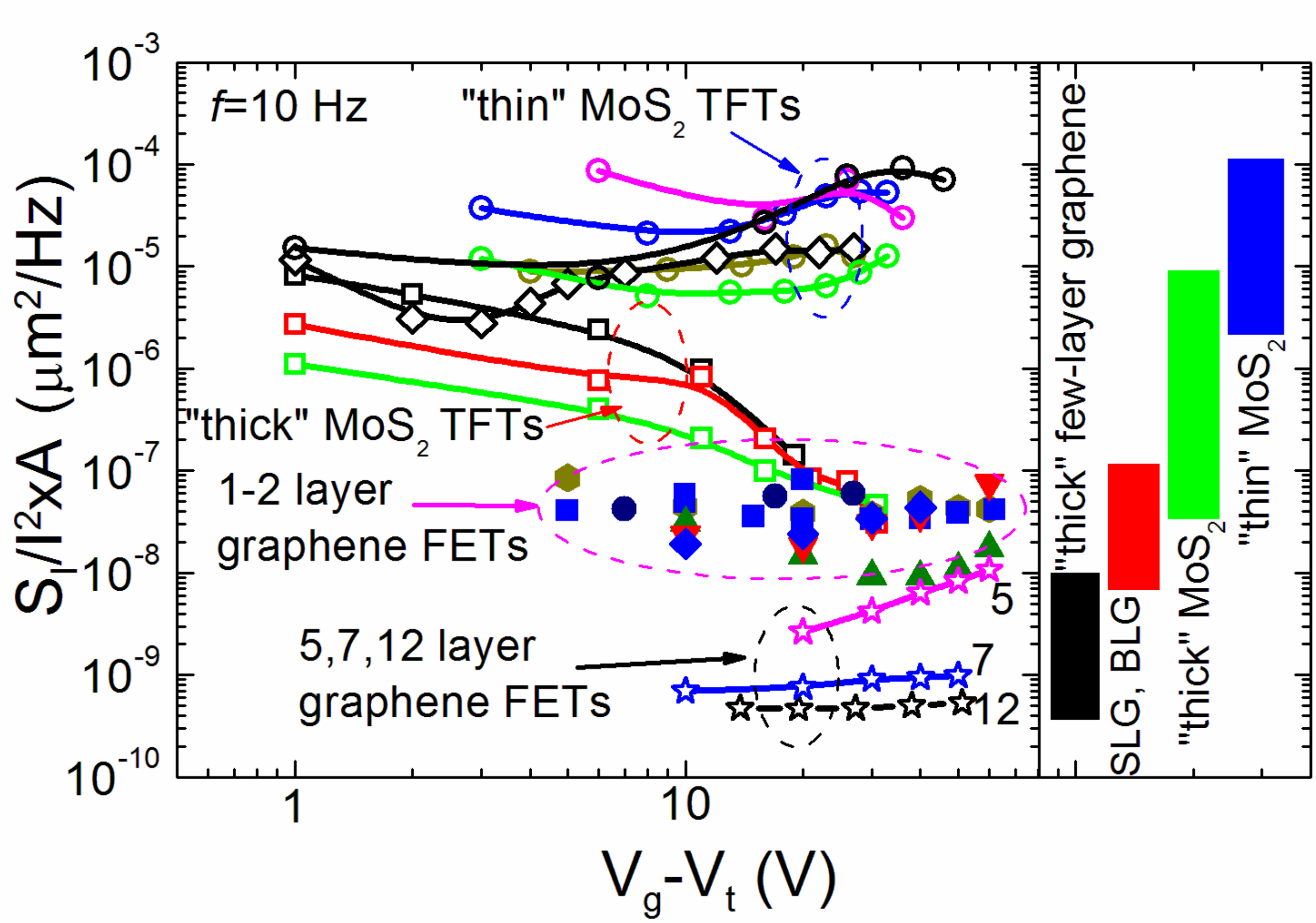
f=10 Hz
"thin" MoS2 TFTs
"thick" MoS2 TFTs
1-2 layer
graphene FETs
5,7,12 layer
graphene FETs
5
7
12
SI/I2xA (μm2/Hz)
Vg-Vt (V)
"thick" few-layer graphene
SLG, BLG
"thick" MoS2
"thin" MoS2

Figure 3